\documentclass[aip,jap,preprint]{revtex4-1}
\usepackage{graphicx}
\draft

\begin{document}

\title{Measurement of brightness temperature of two-dimensional electron gas in channel of a high electron mobility transistor at ultralow dissipation power}

\author{A.M. Korolev}
\email{korol.rian@gmail.com}
\affiliation{Institute of Radio Astronomy, NAS of Ukraine, Chervonopraporna St. 4, Kharkov 61002, Ukraine} %
\author{V.M. Shulga}
\affiliation{Institute of Radio Astronomy, NAS of Ukraine, Chervonopraporna St. 4, Kharkov 61002, Ukraine} %
\author{O.G. Turutanov}
\affiliation{B. Verkin Institute for Low Temperature Physics and Engineering, NAS of Ukraine, Lenin Ave. 47, Kharkov 61103, Ukraine} %
\author{V.I. Shnyrkov}
\affiliation{B. Verkin Institute for Low Temperature Physics and Engineering, NAS of Ukraine, Lenin Ave. 47, Kharkov 61103, Ukraine} %

\date{\today}

\begin{abstract}
A technically simple and physically clear method is suggested for
the direct measurement of brightness temperature of two-dimensional
electron gas (2DEG) in the channel of a high electron mobility
transistor (HEMT). The usage of the method was demonstrated with the
pseudomorphic HEMT as a specimen. The optimal HEMT dc regime, from
the point of view of the "back action" problem, was found to belong
to the unsaturated area of the static characteristics possibly
corresponding to the ballistic electron transport mode. The proposed
method is believed to be a convenient tool to explore the ballistic
transport, electron diffusion, 2DEG properties and other
electrophysical processes in the heterostructures.
\end{abstract}


\maketitle

\section{Introduction}

      It is a common knowledge that the impact of a measuring device
onto the object-under-test should be minimized if working with
signal sources of essentially quantum nature. This is a general
problem of non-disturbing quantum measurements. For electronic
detecting facilities, especially amplifiers, this means minimizing
the energy, of noise or other origin, irradiated backwards to the
object-under-test. The effect is a so called "back action", the
phenomenon causing uncontrolled destruction of a quantum state of
the object, e.g., the qubit decoherencing, etc.
\cite{1Bladh,2Grajcar,3Xue}. The back action is detected and
requires a quantitative description in a wide frequency band, orders
of magnitude wider than the amplifier operation frequency band.
Therefore, the "equivalent noise temperature" ($T_{n}$) determined
for a relatively narrow operation frequency band of the amplifier
(receiver) is an inadequate term here. Instead, one should say about
wide-spectrum brightness temperature of the amplifier and its active
elements, at both the input and output "terminals".

      The amplifiers intended for ultra-low temperature applications
(to amplify the signals from quantum detectors, single electron
transistors and variety of other quantum structures) are typically
based on the field-effect transistors. Among them, a class of HEMT
is distinguished, the field-effect transistors with high electron
mobility. HEMTs feature a very wide operational frequency band while
field-induced as opposed to thermally-generated current-carrier
electrons (two-dimensional electron gas, 2DEG) in the channel
principally enable the transistor functionality down to the absolute
zero temperature. Owing these advantages of the HEMTs they are
widely used in ultra-sensitive readout
amplifiers\cite{4Oukhanski,5DeFeo} for quantum devices signals.
Consequently, the quantitative description of the back action as
applied to HEMTs is a hot issue.

      Thermal noise is generated in the HEMT input (gate-source)
terminals due to power dissipation in the input circuit of the
transistor. The dissipative losses come mainly from the gate
metallization resistance, the under-gate channel resistance and the
source resistance. The corresponding irradiation (for perfect
matching, or zero input reflection coefficient) is characterized by
the gate temperature $T_{g} $ which is close to the physical
temperature of the transistor crystal lattice $T_{lat}$. Cooling
down to the cryogenic temperatures is an effective method to
suppress the thermal irradiation. If an ultra-deep cooling is
supposed, it should be accompanied by a considerable decrease (down
to a few microwatts and less) in the transistor consumed/dissipated
power to avoid excessive Joule overheating of its active area. The
situation with overheating becomes more severe because of low
thermal conductivity of the heterostructures \cite{6Bautista}. The
low cooling capacity of the ultra-low-temperature cryorefrigerators,
especially below 100 mK, also strongly limits the HEMT dissipated
power. Provided if the heat sink is effective, the
input-circuit-generated thermal noise is reduced sufficiently and
can be neglected regarding the back action.

      The "hot" electrons irradiation from the gate-drain channel
region, being another cause of the back action, also contributes
substantially to  $T_{n}$. A high effective electron temperature
$T_{d}$  (drain temperature) exceeding the lattice temperature by
two orders of magnitude is inherent for this mechanism. The
irradiation of the 2DEG in gate-drain part of the channel goes
backward to the input via intrinsic drain-gate capacitance of the
transistor. The excitation of the waveguide modes by the output
circuit in the conductive cavity (where the amplifier is placed and
often the signal source) is an additional way.

      The effect of  $T_{d}$  can be roughly estimated using the
reverse transmission gain ($S_{12}$) from the transistor S-matrix.
Typically,  $S_{12}$  is about -20...-30 dB at 1 GHz frequency. The
$S_{12}$  rises almost linearly with frequency so the effect of
$T_{d}$  can prevail over  $T_{g}$. If the amplifier has a high
input impedance ($S_{12}$  is defined for 50-$\Omega$ network), then
the reverse transmission increases stimulating the search for the
ways of  $T_{d}$  reduction.

      The  $T_{d}$  and  $T_{g}$  are used to calculate basic
noise characteristics of a transistor, namely, the minimal noise
temperature, the optimal source impedance and the noise conductivity
\cite{7Pospieszalski}. The both  $T_{d}$  and  $T_{g}$  figures are
extracted from a series of the noise measurements by solving the
inverse problem \cite{8Stenarson} on the basis of the
electrophysical transistor model which is inevitably limited to
certain frequency band and temperature range. Integrally, the
extraction procedure is sophisticated and ambiguous.

      In the context of the back-action problem, the characteristics
of the existing energy flows disturbing the quantum object should be
found. The temperature is a natural quantifier of the chaotic
(noise) irradiation. So, the task is rather to measure directly the
brightness temperature than to solve the inverse problem on
extraction of  $T_{d}$,  $T_{g}$  or other similar noise invariants
\cite{9Dambrine}.

      First,  to measure $T_d$ instrumentally, the contribution of
the amplified noise of the input circuit to the integral output
noise irradiation of the transistor should be excluded. Referring to
the modern transistors with cut-off frequencies of tens and more
gigahertz, such an elimination is hard enough because of stability
problem. Moreover, it becomes much more complicated under the
deep-cooling conditions. However, the ultra-low power consumption of
the transistor associated with deep-cooled amplifiers results in
decrease in the cut-off frequency by two orders of magnitude while
the stability factor exceeds the unity. Consequently, the stability
is not further an issue, and the direct instrumental measurement of
$T_d$ become possible.

      In this work we propose a simple method to measure directly
the brightness temperature of the 2DEG in a HEMT channel. The
results of the measurements are discussed and the recommendations on
the choice of the HEMT dc regime are formulated concerning the back
action phenomenon.

\section{The brightness temperature measurement technique}

\begin{figure}[h!]
\centering
\includegraphics[width = 0.8\columnwidth]{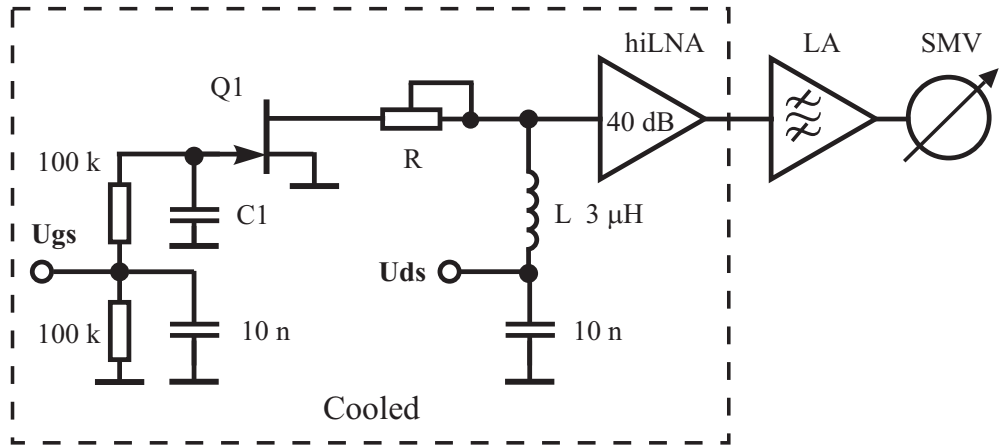}
\caption{\label{fig01}%
The simplified diagram of the testing setup.}
\end{figure}

      The simplified diagram of the experimental setup is shown in
Fig.\,\ref{fig01}. The essence of the technique are by-turn
measurement and subsequent comparison of the powers of the two
signals. The first one is produced by the output circuit of the
transistor-under-test (Q1), the second one is the reference, taken
from a variable resistor R whose temperature is equal to the ambient
temperature. During the calibration procedure, the resistance R is
set equal to the differential resistance of the channel in a
specified point of the transistor static characteristics. The
measurement cycle is described below in more detail.

      The transistor gate is ac-shorted to the source by C1 to
exclude the amplified input circuit noise from the net output
signal.

      The self inductance of the capacitor C1 (SMD 0603, 330 pF)
along with the inductance of the Q1 gate terminal does not exceed 3
nH. The associated reactance at a mean frequency (50 MHz) of the
operational range is inductive and not greater than 4$\;\Omega$ and
the capacitive reactance of the gate-source of more than 3$\;$
k$\Omega$. Under these conditions, only the noise component
representing the source resistance noise is amplified by the
transistor adding to $T_d$. This resistance does not exceed 3$\;
\Omega$ for practically all low-power HEMTs. Taking into account
that the transistor voltage gain does not exceed 5 under the
microcurrent dc regime, it is easy to show that the contribution of
the source noise in the output signal is less than 1\%.

      The transistor Q1 and the variable resistor R are placed close
to the instrumental amplifier to minimize the shunt capacitance. A
three-stage HEMT (AVAGO ATF35143) high input impedance (100$\;$
k$\Omega$), low-noise cooled amplifier (hiLNA in Fig.\,\ref{fig01})
is used as the instrumental amplifier. The hiLNA gain is about 40
dB, the operational frequency band is 20 to 100 MHz. The integral
noise temperatures of the hiLNA are 2.3$\pm$0.5 K and 1.2$\pm$0.5 K
at the source resistance of 10$\;\Omega$ and 1000$\;\Omega$,
respectively, and the ambient temperature  $T_{amb} =4.2\;$K. The
amplifier circuitry is similar to the earlier reported devices
\cite{10Oukhanski}.

      The measuring circuit is a low-Q parallel tank consisted of
the resistor R and HEMT channel resistance, the capacitance of the
wires, pads and hiLNA input (totally 5 pF) and the inductance L
(3$\;\mu$H).

      A linear amplifier (LA) with a band-pass filter
(40...80 MHz) is placed next to the hiLNA. The specified frequency
band (namely, the lower edge) is chosen in order to exclude the
$1/f$ noise of Q1.

      The output signal level is measured by a square-meter
voltmeter (SMV) within an accuracy of better than 1\%. The
multisection filters are put into the supply circuits (not shown in
Fig.\,\ref{fig01}) and the test unit is electromagnetically
shielded. The measurement errors are 0.1\% and 1.5\% for the dc
voltages and currents, correspondingly.

      The measuring cycle is a three-stage one. First, the resistor
R is set to zero resistance, and the static characteristics of the
transistor are measured. The drain-source resistance  $r_{d}$  is
calculated from the data obtained as a function of  $U_{ds}$  and
$U_{gs}$. The SMV readouts are taken for each calculated $r_d$.
According to the measurement procedure and Nyquist theorem, the SMV
voltage, which is proportional to the output signal power, can be
written as:

\begin{equation}\label{1}
\displaystyle \left\langle U_{m}^{2} \right\rangle =\left\langle
U_{0}^{2} \right\rangle +4kr_dT_{el} G_m \Delta F_m,
\end{equation}

\noindent
where $\left\langle U_{m}^{2} \right\rangle$ is the mean
square of the measured output voltage at a specified drain-source
resistance $r_d$, $\left\langle U_{0}^{2} \right\rangle$ is the mean
square of the measured output voltage at $r_c$ = 0 (see below), $k$
is the Boltzmann constant, $T_{el}$ is the brightness temperature of
the noise irradiation of the transistor channel, $G_m$ and $\Delta
F_m$ are the total gain and effective pass band of the measuring
system, correspondingly, at the specified $r_d$.

      The second stage is a calibration. The transistor is
zero-biased and maximally opened ($U_{ds}=U_{gs}=0$). The channel
resistance $r_d$ is minimal ($r_{dmin}$=8$\pm$1 $\Omega$ for
ATF36077) and fully linear. The noise generated by $r_d$ is purely
thermal (white). It adds to the noise of the calibration resistor R
with resistance $r_c$. The calibration includes tuning the
calibration resistor R in the range of 0...670 $\Omega$ while the
SMV readout is synchronously taken. In a manner similar to
(\ref{1}), we write the equation for the SMV output signal:

\begin{equation}\label{2}
\displaystyle \left\langle U_{c}^{2} \right\rangle =\left\langle
U_{0}^{2} \right\rangle +4kr_c T_{amb} G_c \Delta F_c,
\end{equation}

\noindent where  $\left\langle U_{c}^{2} \right\rangle $  is the
mean square of the measured output voltage at a specified resistance
$r_c$ of the calibration resistor R (exactly, $r_{dmin}+r_c$),
$T_{amb}$  is the physical temperature of the calibration resistor
R, $G_c$ and $\Delta F_c$ are the total gain and the effective pass
band of the measuring system, correspondingly, at the specified
$r_c$.

Finally, an expression for $T_{el}$ can be derived from (\ref{1})
and (\ref{2}) taking $G_c=G_m$, $\Delta F_c=\Delta F_m$. These
equalities are valid if the experimental values are taken from the
data array with the selection rule $r_d=r_c$. Then we obtain:

\begin{equation}\label{3}
\displaystyle T_{el}=T_{amb} \frac{\left\langle U_m^{2}
\right\rangle -\left\langle U_0^{2} \right\rangle}{\left\langle
U_c^{2} \right\rangle -\left\langle U_0^{2} \right\rangle}
\end{equation}

To simplify the measurement and calculation procedures, a few
assumptions were made when writing (\ref{1})-(\ref{3}).

(i) $\left\langle U_0^{2} \right\rangle$=const is assumed, i.e. the
noise temperature of the instrumental amplifier (hiLNA) does not
depend on the source resistance. Actually, it varies (see above)
although always staying below the temperature to measure, $T_{el}$.

(ii) $\left\langle U_0^{2} \right\rangle$ is supposed to be measured
at $r_c$=0.

In fact, the minimal value $r_{cmin}= r_{dmin} \approx$ 8
$\Omega$ (for ATF36077).

(iii) It is believed that the dc channel resistance $r_d$ is equal
to the ac one, at the measurement frequency (40-60 MHz). In reality,
the ac channel resistance is less by a few percent.

All the assumptions made are not too rough, and the systematic
absolute error (see section III) does not exceed 2 K. Nevertheless,
it should be taken into account when $T_{el}$ is about several
kelvins. When the measured $T_{el}$ is of some tens kelvins (the
most important and interesting range), the error is determined by
the accuracy of the instrumental measurement being, by our estimate,
of about 10\% of the measured value. To the authors' opinion, this
error is acceptable in the context of this work.

      Additionally, the following should be clarified here. The
drain temperature $T_{d}$, as a parameter in the two-temperature
Pospeshalski noise model \cite{7Pospieszalski}, actually represents
the effective temperature of the 2DEG in the saturated region
(roughly, the gate-drain region) of the HEMT channel. Meanwhile, the
expression (\ref{3}) that we obtained for  $T_{el}$ represents an
averaged 2DEG temperature throughout the whole channel including the
source-gate part with its temperature which is always close to
$T_{lat}$. It is the 2DEG temperature that we designate by $T_{el}$.
It is measured directly and has a clear physical sense. However,
$T_{d}$  and $T_{el}$ do not differ too much since the resistance of
the source-gate region is about 1/100 of the total channel
resistance.

      The measurement of  $T_{el}$  can, in principle, be done
without the calibration resistor, using the transistor noise instead
at  $U_{ds}$ =0,  $I_{d}$ =0. However, in practice this requires an
extra (negative) bias voltage which could cause, for some
transistors, the gate leakage noise to emerge.

\section{Brightness temperature measurement results and discussion}

      The pseudomorphic HEMT AVAGO ATF36077 was chosen as a test
object. We believe this transistor is a good representative of
low-power HEMTs. Additionally, as our practice show, ATF36077 has a
very low gate leakage current and a high mechanical stability during
multiple thermal cycling. The ATF36077 static characteristics show
no hysteresis. All said above makes ATF36077 a suitable object for
testing the proposed measurement method.

      The bias voltage and, correspondingly, the maximal drain
current were chosen so that the transistor dissipated power  $P_{c}$
would not exceed $200\; \mu$W. The larger  $P_{c}$  values are
beyond the subject of this work. Also, at  $P_{c} >50\; \mu$W the
transistor-under-test should be carefully monitored for a parasitic
oscillation to emerge which is provoked by the gate grounding. The
consumption current of the hiLNA is a suitable indicator of the
parasitic oscillations.

      The measurements were taken at 4.2 K, 78 K and 290 K when
developing the technique. The results of the measurements at 4.2 K
are reported below.

\begin{figure}[h!]
\centering
\includegraphics[width = 0.8\columnwidth]{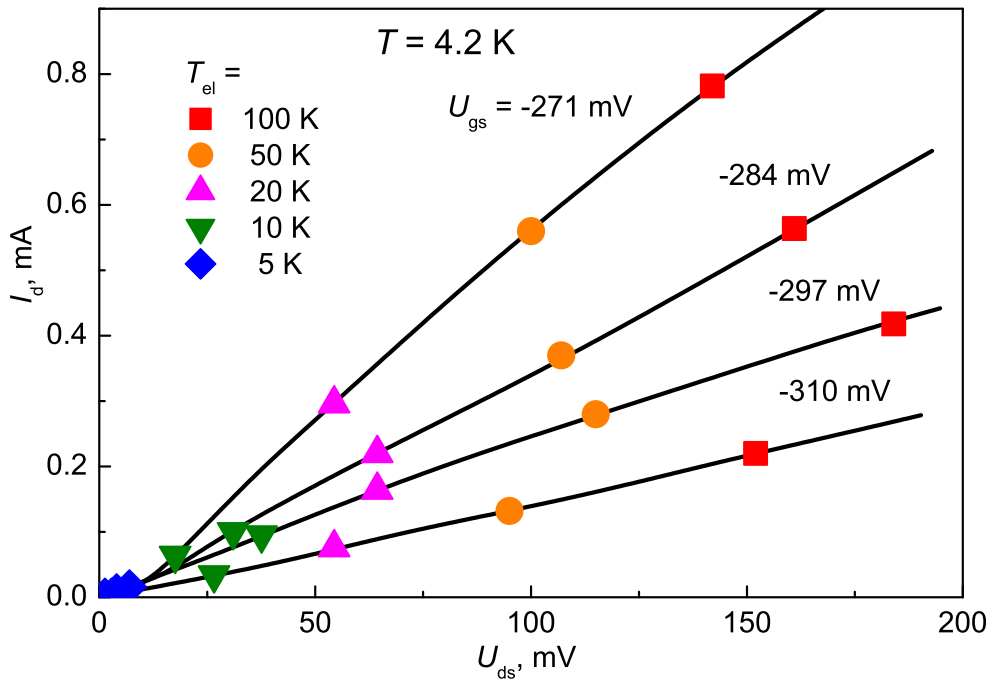}
\caption{ \label{fig02} %
(Color online) The static I-V characteristics of ATF36077 and the
electron temperature at 4.2 K.}
\end{figure}

      Fig.\,\ref{fig02} demonstrates the static characteristics and
the electron temperature for ATF36077. Let us start with these data
to estimate the measurement error. To do this, we note that the
measured $T_{el}$ is assumed to be equal to $T_{amb}$ when $U_{ds}$
= 0, $I_d$ = 0. It is quite reasonable from the physical point of
view. Then the difference between $T_{el}$ and $T_{amb}$ in the
origin of the static characteristic branch ($U_{ds}$=0) can serve as
a measure of the absolute error. The data show that the systematic
error does not exceed 2 K.

Let us take conventionally  $S_{12} =-10\;$ dB  as a measure of the
back action. Also, we define the criterion of the acceptable
irradiation in the direction of the signal source: the temperature
of such an irradiation must not be higher than $2T_{lat}$, that is
$T_{el}\leq 20 T_{amb}$. Bearing in mind these definitions, a
recommendation useful in designing the amplifiers with reduced back
action can be derived from Fig.\,\ref{fig02}. Namely, the maximal
acceptable value of $U_{ds}$  is 150 mV for HEMT operating at
$T_{amb}=4.2$ K.

\begin{figure}[h!]
\centering
\includegraphics[width = 0.8\columnwidth]{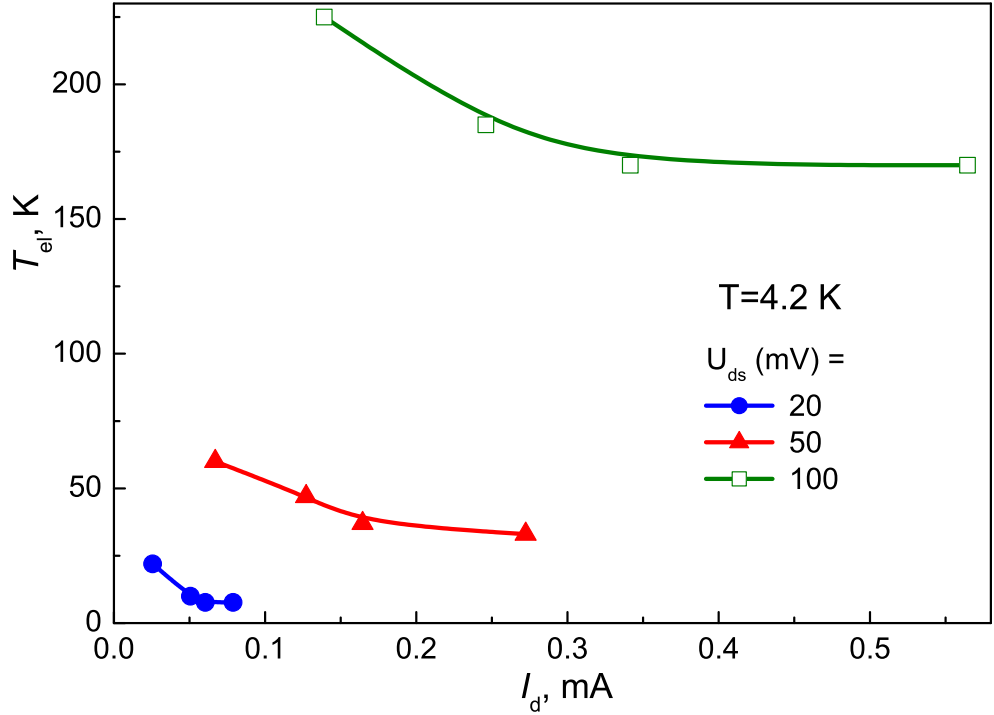}
\caption{ \label{fig03} %
(Color online) The electron temperature as a function of the drain
current at fixed drain-source voltages.}
\end{figure}

      Fig.\,\ref{fig03} shows the electron temperature  $T_{el}$  as a
function of the drain current  $I_{d}$  at fixed drain-source
voltages  $U_{ds}$.

      It is clearly seen that  $T_{el} (I_{d})$  plot tends to a
saturation and, there is no increase of the electron temperature
with the current anyway. This witnesses for a negligible
contribution of the channel diffusion noise induced onto the gate.
Therefore, the gate is "well-grounded" while the operation frequency
and the circuit design are adequate.

\begin{figure}[h!]
\centering
\includegraphics[width = 0.8\columnwidth]{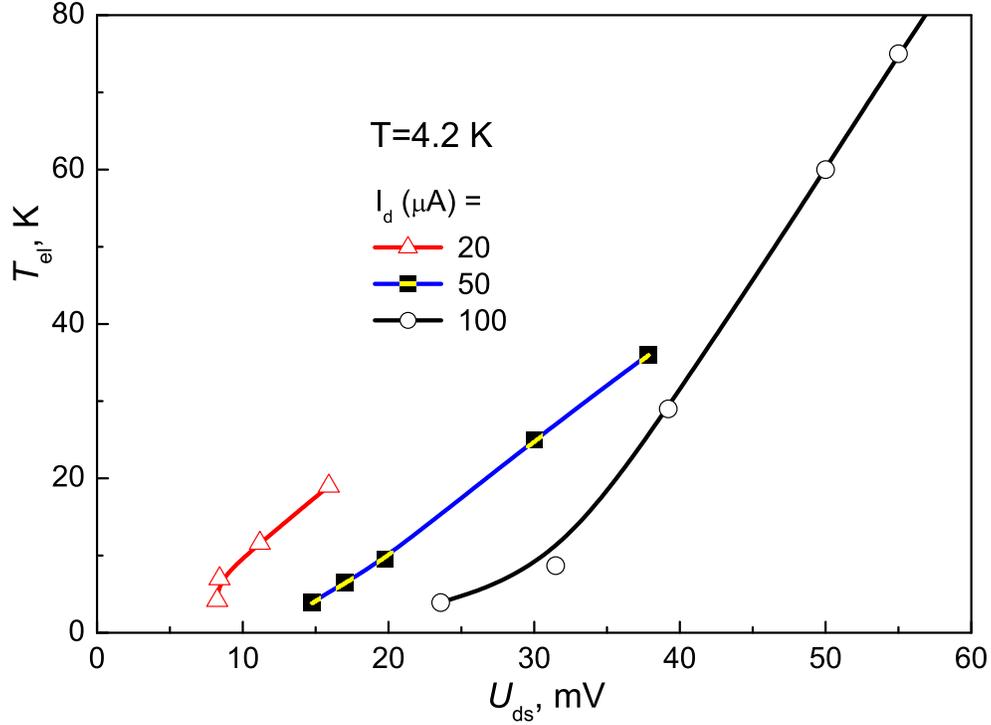}
\caption{\label{fig04} %
(Color online) The electron temperature as a function of the
drain-source voltage at fixed drain currents.}
\end{figure}

      The  $T_{el}$ as a function of  $U_{ds}$  at fixed drain
currents is plotted in Fig.\,\ref{fig04}. It should be noted at once
that the doubled electron temperature ($T_{el} =2T_{amb}$) is
observed at the voltages much higher than expected threshold values
(of order of 1 mV) approximately following from the condition
$kT=eU_{ds}$. The effect could evidence for the ballistic
(collision-free) electron transport at corresponding parts of the
static characteristics.

      A distinctive bend of the curve taken at the drain current of
100 $\mu$A may be associated with the threshold of the optical
phonon scattering mechanism activation (approximately 30 eV for the
A3B5 semiconductors). The corresponding dc regime ($I_{d} \approx
0.1\;$mA,  $U_{ds} \le 35\;$mV) can be recommended for the
first-stage transistors in the amplifiers with minimized back action
intended to function at subkelvin temperatures. The HEMT ability of
working at ultra-low supply voltages in the unsaturated region of
the static characteristics was reported earlier in
\cite{12Korolev,13Korolev}. The absence of the "bend" in the plots
corresponding to the drain currents 20 and 50 $\mu$A is most likely
due to the expansion of the under-gate depletion region towards the
source. This effect probably causes the paradoxical rise of $T_{el}$
with decreasing the channel current (Fig.\,\ref{fig04}) as well.

The above estimates and the plots analysis is rather qualitative.
They aimed, in the context of this work, to the demonstration of
reasonability of the measurement of the electron temperature in
studying electrophysical processes in HEMTs and finding the optimal
regimes from the point of view of the back action problem.

      To estimate the back action at low frequencies where
$1/f$  noise predominates, it is necessary to measure the amplifier
noise temperature with the standard procedure \cite{4Oukhanski}.

\section{Summary}

      A technique is suggested in this paper for the direct
measurement of the brightness temperature of the 2DEG in a HEMT. The
technical simplicity and clear physical sense of the measurement
results makes the proposed method a convenient tool for further
studies of the delicate electrophysical processes such as the
ballistic electron transport and diffusion, the 2DEG properties and
so on.

      Grounding on the results of the 2DEG brightness temperature
measurements, the recommendations are put for choosing the HEMT dc
regimes which are the best concerning the back action. It is found
that the optimal HEMT operation points to minimize the back action
lie in the unsaturated area of the static characteristics that
possibly corresponds to the electron ballistic transit.


\end{document}